# INCREASED PREDICTION ACCURACY IN THE GAME OF CRICKET USING MACHINE LEARNING

Kalpdrum Passi and Niravkumar Pandey

Department of Mathematics and Computer Science
Laurentian University, Sudbury, Canada

## ABSTRACT

*Player selection is one the most important tasks for any sport and cricket is no exception. The performance of the players depends on various factors such as the opposition team, the venue, his current form etc. The team management, the coach and the captain select 11 players for each match from a squad of 15 to 20 players. They analyze different characteristics and the statistics of the players to select the best playing 11 for each match. Each batsman contributes by scoring maximum runs possible and each bowler contributes by taking maximum wickets and conceding minimum runs. This paper attempts to predict the performance of players as how many runs will each batsman score and how many wickets will each bowler take for both the teams. Both the problems are targeted as classification problems where number of runs and number of wickets are classified in different ranges. We used naïve bayes, random forest, multiclass SVM and decision tree classifiers to generate the prediction models for both the problems. Random Forest classifier was found to be the most accurate for both the problems.*

## KEYWORDS

*Naïve Bayes, Random Forest, Multiclass SVM, Decision Trees, Cricket*

## 1. INTRODUCTION

Cricket is a sport played by two teams with each side having eleven players. Each team is a right blend of batsmen, bowlers and allrounders. The batsmen's role is to score maximum runs possible and the bowlers have to take maximum wickets and restrict the other team from scoring runs at the same time. Allrounders are the players who can both bat and bowl and they contribute by scoring runs and taking wickets. Each player contributes towards the overall performance of the team by giving his best performance in each match. Each player's performance varies with factors like the team he is playing against and the ground at which the match is being played. It is important to select the right players that can perform the best in each match. The performance of a player also depends on several factors like his current form, his performance against a particular team, his performance at a particular venue etc. The team management, the coach and the captain analyze each player's characteristics, abilities and past stats to select the best playing XI for a given match. In other words, they try to predict the players' performance for each match.

In this paper, we predict the players' performance in One Day International (ODI) matches by analyzing their characteristics and stats using supervised machine learning techniques. For this, we predict batsmen's and bowlers' performance separately as how many runs will a batsman score and how many wickets will a bowler take in a particular match.

## 2. LITERATURE REVIEW

An extensive online search produced very few articles related to players' performance prediction in the game of cricket. A very small number of researchers have studied the performance of cricket players. Muthuswamy and Lam[1] predicted the performance of Indian bowlers against seven international teams against which the Indian cricket team plays most frequently. They used back propagation network and radial basis network function to predict how many runs a bowler is likely to concede and how many wickets a bowler is likely to take in a given ODI match. Wikramasinghe[2] predicted the performance of batsmen in a test series using a hierarchical linear model. Barr and Kantor[3] defined a criterion for comparing and selecting batsmen in limited overs cricket. They defined a new measure P(out) i.e. probability of getting out and used a two-dimensional graphical representation with Strike Rate on one axis and P(out) on another. Then they define a selection criterion based on P(out), strike rate and batting average of the batsmen. Iyer and Sharda[4] used neural networks to predict the performance of players where they classify batsmen and bowlers separately in three categories – performer, moderate and failure. Based on the number of times a player has received different ratings, they recommend if the player should be included in the team to play World Cup 2007. Jhanwar and Paudi[5] predict the outcome of a cricket match by comparing the strengths of the two teams. For this, they measured the performances of individual players of each team. They developed algorithms to model the performances of batsmen and bowlers where they determine the potential of a player by examining his career performance and then his recent performances. Lemmer[6] defined a new measure called Combined Bowling Rate to measure the performance of bowlers. The combined bowling rate is a combination of three traditional bowling measures: bowling average, strike rate and economy. Bhattacharjee and Pahinkar.[7] used this combined bowling rate to analyze the performance of bowlers in Indian Premier League (IPL). They also determined other factors that affect the performance of bowlers and applied multiple regression model to identify the factors that are empirically responsible for the performance of bowlers. Mukharjee.[8] applied Social Network Analysis to rate batsmen and bowlers in a team performance. He generated a directed and weighted network of batsmen-bowlers using player-vs-player information available for test and ODI cricket. He also generated a network of batsmen and bowlers using the dismissal record of batsmen in the history of cricket. Shah[9] also defined new measures to measure players' performance. The new measure for batsmen takes into account the quality of each bowler he is facing and the new measure for bowlers considers the quality of each batsman he is bowling to. The aggregate of individual performance of a batsman against each bowler is the total performance index of the batsman. Similarly, the aggregate of individual performance of a bowler against each batsman is the total performance index of the bowler. Parker, Burns and Natarajan.[10] defined a model for valuation of players for IPL auction. Their model considered factors like previous bidding price of the player, experience of the player, strike rate etc. Prakash, Patvardhan. and Lakshmi[11] defined batting index and bowling index to rank players' performance for their models to predict outcomes of IPL matches. Ovens and Bukiet [12] applied a mathematical approach to suggest optimal batting orders for ODI matches. Schumaker et. el. [13] described how statistical simulations can be used in predictive modelling for different sports. Haghighat et. el. [14] reviewd the data mining systems previously used in sports prediction and describe advantages and disadvantages of each system. Hucaljuk and Rakipovik [15] used machine learning techniques to predict outcomes of football matches.McCullagh [16] used neural networks for player selection in Australian Footbal League.

Our work is probably the first generalized approach to predict how many runs will a batsman score and how many wickets will a player take on a particular match day. Muthuswamyand Lam[1] carried out a similar study predicting how many wickets will a bowler take using neural networks but their work was limited to eight Indian bowlers and is difficult to generalize for all the bowlers in the world. We used some supervised machine learning algorithms to build prediction models that can be used to predict the performance of any player in a given match.

## 3. DATA AND TOOLS

We obtained all our data from www.cricinfo.com using scraping tools, parsehub[17] and import.io[18]. For batting, we considered matches played from January 14, 2005 to July 10, 2017. The senior most player during this span was SR Tendulkar, so we collected innings by innings list of the performance of all the batsmen from December 18, 1989 when he played his first ODI match. For bowling, we considered matches played from January 2, 2000 to July 10, 2017. The senior most player during this span was PA de Silva, so we collected innings by innings list of the performance of all the batsmen from March 31, 1984 when he played his first ODI match. Since the past stats of the players such as average, strike rate etc. are not available directly online for each match they played, we calculated from the innings by innings list for each match. We imported all the data in MySQL tables and used php to manipulate them.

For predictive analytics, we used Weka and Dataiku. Both these tools are a collection of machine learning algorithms for data mining and also provide some preprocessing functionalities. All the results in this study have been obtained from Weka 3-9-1-oracle-jvm and Dataiku Data Science Studio on Mac OS 10.11.6 and Windows 10.

## 4. DATA PREPROCESSING

### 4.1 CALCULATING THE TRADITIONAL ATTRIBUTES

As mentioned in the previous section, the stats of the players such as average, strike rate etc. are not available directly for each game, we calculated these attributes from the innings by innings list using aggregate functions and mathematical formulae. These attributes are generally used to measure a player's performance. These attributes are as follows:

### 4.1.1 Batting Attributes

**No. of Innings:** The number of innings in which the batsman has batted till the day of the match. This attribute signifies the experience of the batsman. The more innings the batsman has played, the more experienced the player is.

**Batting Average:** Batting average commonly referred to as average is the average number of runs scored per innings. This attribute indicates the run scoring capability of the player.

**Average = Runs Scored / Number of times dismissed**

**Strike Rate (SR):** Strike rate is the average number of runs scored per 100 balls faced. In limited overs cricket, it is important to score runs at a fast pace. More runs scored at a slow pace is rather harmful to the team as they have a limited number of overs. This attribute indicates how quickly the batsman can score runs.

**Strike Rate: (Runs Scored / Balls Faced) * 100**

**Centuries:** Number of innings in which the batsman scored more than 100 runs. This attribute indicates the capability of the player to play longer innings and score more runs.
Fifties: Number of innings in which the batsman scored more than 50 (and less than 100) runs.This attribute indicates the capability of the player to play longer innings and score more runs.

**Zeros:** Number of innings in which the batsman was dismissed without scoring a single run. This attribute shows how many times the batsman failed to score runs, hence this being a negative factor, was impacts the batsman's prediction negatively.

**Highest Score:** The highest runs scored by a batsman in any (single) innings throughout his career. This attribute is used in the formula for calculating the venue attribute. This attribute shows the run scoring capability of the batsman at the venue. If a player has s very high score at a venue in past, he is more likely to score more runs at that venue.

### 4.1.2 Bowling Attributes

**No. of Innings:** The number of innings in which the bowler bowled at least one ball. It represents the bowling experience of a player. The more innings the player has played, the more experienced the player is.

**Overs:** The number of overs bowled by a bowler. This attribute also indicates the experience of the bowler. The more overs the bowler has bowled, the more experienced the bowler is.

**Bowling Average:** Bowling average is the number of runs conceded by a bowler per wicket taken. This attribute indicates the capabilities of the bowler to restrict the batsmen from scoring runs and taking wickets at the same time. Lower values of bowling average indicate more capabilities.

**Bowling Average: Number of runs conceded / Number of wickets taken**

**Bowling Strike Rate:** Bowling strike rate is the number of balls bowled per wicket taken. This attribute indicates the wicket taking capability of the bowler. Lower values mean that the bowler is capable of taking wickets quickly.

**Strike Rate: Number of balls bowled / Number of wickets taken**

**Four/Five Wicket Haul:** Number of innings in which the bowler has taken more than four wickets. This attribute indicates the capability of the bowler to take more wickets in an innings. Higher the value, more capable the player.

### 4.2 CALCULATING THE WEIGHTS

As we saw, different measures signify different aspects of a player's abilities and hence some measures have more importance than others, e.g. batting average is an important factor for all the formats of the game as it reflects the run scoring abilities of a batsman in general. Similarly, strike rate would be an important factor for limited over matches as it is important to score more runs in limited overs. So, we weighted each measure of performance according to its relative importance over other measures. We determined the weights using analytic hierarchy process(AHP)[19][20]. AHP is an effective tool for complex decision making. It aids in setting priorities and making the best decision. AHP reduces complex decisions into a series of pairwise comparisons. AHP captures both subjective and objective aspects of a decision. The AHP generates a weight for each evaluation criterion according to the decision maker's pairwise comparisons of the criteria. The higher the weight, the more important the corresponding criterion. Next, for a fixed criterion, the AHP assigns a score to each option according to the decision maker's pairwise comparisons of the options based on that criterion. The higher the score, the better the performance of the option with respect to the considered criterion. Finally, the AHP combines the criteria weights and the options scores, thus determining a global score for each option, and a consequent ranking. The global score for a given option is a weighted sum of the scores it obtained with respect to all the criteria.

### 4.3 CALCULATING THE DERIVED ATTRIBUTES

To predict a player's performance, his past performances need to be analyzed in terms of how much experience does he have, how consistent he has been in his performance, how well he has been performing in recent matches, how well can he tackle the bowlers/batsmen of different teams, how well does he play at different venues, etc. Traditional measures of players' performance cannot reflect these factors directly. So, we tried to reflect and quantify them by deriving four new measures from the traditional measures. These attributes are weighted averages of the traditional attributes. These attributes are explained as follows:

### 4.3.1 Consistency

This attribute describes how experienced the player is and how consistent he has been throughout his career. All the traditional attributes used in this formula are calculated over the entire career of the player.

**Formula for batting:**

$$\text{Consistency} = 0.4262*\text{average} + 0.2566*\text{no. of innings} + 0.1510*\text{SR} + 0.0787*\text{Centuries} + 0.0556*\text{Fifties} - 0.0328*\text{Zeros}$$

**Formula for bowling:**

$$\text{Consistency} = 0.4174*\text{no. of overs} + 0.2634*\text{no. of innings} + 0.1602*\text{SR} + 0.0975*\text{average} + 0.0615*\text{FF}$$

### 4.3.2 Form

Form of a player describes his performance over last one year. All the traditional attributes used in this formula are calculated over the matches played by the player in last 12 months from the day of the match.

**Formula for batting:**

$$\text{Form} = 0.4262*\text{average} + 0.2566*\text{no. of innings} + 0.1510*\text{SR} + 0.0787*\text{Centuries} + 0.0556*\text{Fifties} - 0.0328*\text{Zeros}$$

**Formula for bowling:**

$$\text{Form} = 0.3269*\text{no. of overs} + 0.2846*\text{no. of innings} + 0.1877*\text{SR} + 0.1210*\text{average} + 0.0798*\text{FF}$$

### 4.3.3 Opposition

Opposition describes a player's performance against a particular team. All the traditional attributes used in this formula are calculated over all the matches played by the player against the opposition team in his entire career till the day of the match.

**Formula for batting:**

$$\text{Opposition} = 0.4262*\text{average} + 0.2566*\text{no. of innings} + 0.1510*\text{SR} + 0.0787*\text{Centuries} + 0.0556*\text{Fifties} - 0.0328*\text{Zeros}$$

**Formula for bowling:**

$$\text{Opposition} = 0.3177*\text{no. of overs} + 0.3177*\text{no. of innings} + 0.1933*\text{SR} + 0.1465*\text{average} + 0.0943*\text{FF}$$

### 4.3.4 Venue

Venue describes a player's performance at a particular venue. All the traditional attributes used in this formula are calculated over all the matches played by the player at the venue in his entire career till the day of the match.

**Formula for batting:**

Venue = 0.4262*average + 0.2566*no. of innings + 0.1510*SR + 0.0787*Centuries + 0.0556*Fifties + 0.0328*HS

**Formula for bowling:**

Venue = 0.3018*no. of overs + 0.2783*no. of innings + 0.1836*SR + 0.1391*average + 0.0972*FF

### 4.3.5 Rating The Attributes

The values of the traditional attributes fall in very wide ranges and small differences in these values do not discriminate different players, e.g. batsmen having batting averages of 32.00, 35.50 and 38.60 are considered to be of same quality. So, we rated each traditional measure from 1 to 5 based on the range in which its value falls,to calculate the derived attributes, with 1 being the minimum and 5 being the maximum. We looked at the values of these attributes for different players and applied our knowledge to rate the measures, e.g. some of the best batsmen of the world have had batting averages greater than or equal to 40 for most of the time during their career and generally, averages greater than or equal to 40 are considered excellent, so we rated such batsmen 5 for averages greater than 39.99. We used these ratings instead of actual values of the measures, in the formulae of derived attributes. The measures are rated as follows:

No. of Innings:

For Consistency:
1 – 49: 1
50 – 99: 2
100 – 124: 3
125 – 149: 4
>=150: 5

For Form:
1 – 4: 1
5 – 9: 2
10 – 11: 3
12 – 14: 4
>=15: 5

For Opposition:
1 – 2: 1
3 – 4: 2
5 – 6: 3
7 – 9: 4
>=10: 5

For Venue:
1: 1
2: 2
3: 3
4: 4
>=5: 5

Batting Average (for all derived attributes):

0.0  - 9.99: 1
10.00 - 19.99: 2
20.00 - 29.99: 3
30.00 - 39.99: 4

>=40: 5

Batting Strike Rate (for all derived attributes):

0.0   - 49.99: 1
50.00 - 59.99: 2
60.00 - 79.99: 3
80.00 - 100.00: 4
>=100.00: 5

Centuries:

For Consistency:
1 – 4: 1
5 – 9: 2
10 – 14: 3
15 – 19: 4
>=20: 5

For Form:
1: 1
2: 2
3: 3
4: 4
>=5: 5

For Opposition:
1: 3
2: 4
>=3: 5

For Venue:
1: 4
>=2: 5

Fifties:

For Consistency:
1 – 9: 1
10 – 19: 2
20 – 29: 3
30 – 39: 4
>=40: 5

For Form & Opposition:
1 – 2: 1
3 – 4: 2
5 – 6: 3
7 – 9: 4
>=10: 5

For Venue:
1: 4
>=2:– 5

Zeros:

For Consistency:
1 – 4: 1
5 – 9: 2
10 – 14: 3
15 – 19: 4
>=20: 5

For Form & Opposition:

        1: 1

        2: 2

        3: 3

        4: 4

        >=5: 5

Highest Score (For Venue Only):

        $1 - 24: -1$

        $25 - 49: 2$

        $50 - 99: 3$

        $100 - 150: 4$

        >=150: 5

Overs:

    For Consistency:

        $1 - 99: 1$

        $100 - 249: 2$

        $250 - 499: 3$

        $500 - 1000: 4$

        >=1000: 5

    For Form & Opposition:

        $1 - 9: 1$

        $10 - 24: 2$

        $25 - 49: 3$

        $50 - 100: 4$

        >=100: 5

    For Venue:

        $1 - 9: 1$

        $10 - 19: 2$

        $20 - 29: 3$

        $30 - 39: 4$

        >=40: 5

Bowling Average (for all derived attributes):

    0.00 - 24.99: 5

    25.00 - 29.99: 4

    30.00 - 34.99: 3

    35.00 - 49.99: 2

    >=50.00: 1

Bowling Strike Rate (for all derived attributes):

    0.00 - 29.99: 5

    30.00 -39.99: 4

    40.00 -49.99: 3

    50.00 -59.99: 2

    >=60.00: 1

Four/Five Wicket Haul:

    For Consistency:

        $1 - 2: 3$

        $3 - 4: 4$

        >=5: 5

    For Form, Opposition & Venue:

        $1 - 2: 4$

        >=3: 5

### 4.3.6 Other Input Attributes

Our experiments showed that the derived attributes themselves are sufficient to accurately predict players' performance. Also, there are some other factors apart from past performances that affect players' performances, e.g. depending on the types of bowlers the opposition team has, it would be better to include more left-handed batsmen than right-handed batsmen in the team or vice versa. So, we incorporated more attributes which indicate the players', the opponents' and the venues' characteristics, in our experiments. These attributes are explained below:

Batting Hand: The dominant hand of the batsman while batting. It has two possible values: Left or Right.Depending on the characteristics of the bowlers of the opposition team, left-handed batsmen might perform better than the right-handed batsmen or vice versa.

Bowling Hand: The dominant hand of the batsman while bowling. It has two possible values: Left or Right.Depending on the characteristics of the opposition team's batsmen, left-handed bowlers might perform better than the right-handed bowlers or vice-versa.

Batting Position: The number at which the batsman bats in the batting order.Different batsmen tend to play better at certain numbers. So, sending a batsmen at a particular number will make him more comfortable at play, e.g. M S Dhoni has been playing better at position 7 than other positions.

Match Type: The type of the match. This attribute has four possible values: Normal, quarter-final, semi-final or final.Different types of matches have different levels of importance which affects players' performance, e.g. final matches are more important than normal matches. Moreover, different players are more comfortable and have shown better performances in some types of matches, e.g. some players tend to play well in normal matches but fail in semi-finals and finals or vice versa.

Match Time: The time at which the match is played. There are two possible values: Day or Day-night. The time of the match also affects players' performance depending on different factors like weather, visibility, location etc.

Strength of opposition: This is the batting/bowling strength of the opposition team. It is the average of the consistency measure of the batsmen/bowlers of the opposition team.Players find it easy to score runs/take wickets against weaker teams than stronger teams.

Ven: The relative venue for the teams. It has three possible values: Home, Away or Neutral.The relative venue of the match is certainly a factor that affects players' performance. Some players perform better at home while some play better away from home.

Oppo: The opposition team.Players usually tend to perform better against some teams. This attribute also incorporates the characteristics of the opposition team's players in general.
The following attributes are used only for predicting runs.

**Role: The playing role of the player. It can take following values:**

**Opening Batsman (OBT)** – The two batsmen who usually bat at position one or two are called opening batsmen.

**Top Order Batsman (TOB)** – The batsmen who usually bat at position three or five are called top order batsmen.

**Middle Order Batsman (MOB)** - The batsmen who ususally bat at position five to eight are called middle order batsmen.

**Batsman** – The batsmen who usually bat at different positions are categorised simply as batsmen here.

**Allrounder –** The players who are equally skilled at both batting and bowling are called all rounders.

**Batting Allrounder** – The players who can both bat and bowl but are more skilled at batting than bowling, are called batting allrounders.

**Bowling Allrounder** – The players who can both bat and bowl but are more skilled at bowling than batting, are called bowling allrounders.

**Bowler** – The players who are expert bowlers but not so skilled at batting, are categorised as bowlers.

**Captain:** This is a binary attribute indicating whether a player is captain of the team.This attribute tries to indicate the control and responsibilities the player has. Some players perform well as captains while some perform worse.

**WK:** This is a binary attribute indicating whether a player is a wicketkeeper.Wicketkeepers are primarily batsmen. They are expected to score more runs as they specialize in batting and are less fatigued than other players as they are physically less active during fielding compared to other fielders.

**Innings**: This attributes indicates if it is the first or the second innings of the match.Depending on different factors like time of the match, the venue, the characteristics of the pitch, etc., sometimes it is more desirable to bat in the first innings while sometimes it is better to bowl in the first innings.

**Tournament:** The type of tournament in which the match is being played.Players feel different levels of pressure and go through psychological ups and downs during different types of tournaments. This attribute can take following values:

- Two Team Tournament (TT)
- Three-Four Team Tournament (TFT)
- Five Team Tournament (FT)

**Toss:** Indicates whether the player's team won or lost the toss.Toss affects the mental state of the players as winning the toss gives them the power to decide whether to bat first or to bowl first and gives a strategic lead to the team.

**Pressure:** Indicates mental and psychological pressure on the player. It takes values from 1 to 5. Its value depends on the type of match being played and the teams that are playing the match. The values are defined as follows:

Normal matches: 1
Quarter Finals: 3
Semi Finals: 4
Finals: 5

Above values are incremented by 1 if the match is between India and Pakistan or Australia and England as these countries are strong rivals of each other.

**Host:** The country in which the match is being played.Some players tend to perform better in certain countries as shown by their stats. This attribute also tries to incorporate the general nature of the pitches of different grounds in the country, e.g. Australian and South African venues are known to have bouncy pitches which are helpful to pace bowlers whereas pitches in India are usually dry and are more supportive to spin bowlers.

**Ground:** The ground on which the match is being played.The data about different pitches is not available at this time, so we tried to incorporate the general nature of the pitches at different grounds using this attribute. Also, players are more comfortable at some venues, e.g. a player who has had some world records at a particular ground, is more likely to perform better on that ground.

### 4.3.7 Outputs

Both the problems are treated as classification problems.

Runs are predicted in five classes:
  $1 - 24$: 1
  $25 - 49$: 2
  $50 - 74$: 3
  $75 - 99$: 4
  >=100: 5
Wickets are predicted in three classes:
  0 - 1: 1
  $2 - 3$: 2
  >=4: 3

### 4.3.8 Data Cleaning

A large number of values of Opposition and Venue were zero. This is because a player has not played any match against a particular team or at a venue before the day of play. We treated such values as missing values and replaced them with the class average of corresponding attributes.

### 4.3.9 Class Imbalance

We observed that majority of the records fall within class 1 in both batting and bowling. This created a major imbalance in the distribution of values and affected the performance of the learning algorithms. To solve this problem, we applied an oversampling technique Supervised Minority Oversampling Technique (SMOTE) **[21]** on minority classes to make all the classes equally distributed. SMOTE over-samples minority classes by creating synthetic example tuples. To create synthetic tuples of minority class, SMOTE takes each minority class sample and creates synthetic examples along the line segment joining any or all of its nearest neighbors. To generate a synthetic sample, the difference between the feature vector under consideration and its nearest neighbor is taken. This difference is then multiplied by a random number between zero and one and the product is added to the feature vector under consideration. This way, a random point along the line segment joining two specific features is selected. Neighbors from the k nearest neighbors are selected based on the amount of oversampling required. e.g. to oversample a minority class by 300%, three neighbors from a tuple's nearest neighbors are selected and one sample in the direction of each is generated.

## 5. LEARNING ALGORITHMS

For generating the prediction models, we used supervised machine learning algorithms. In supervised learning algorithms, each training tuple is labeled with the class to which it belongs[22]. We used naïve bayes, decision trees, random forest and multiclass support vector machines for our experiments. These algorithms are explained in brief.

### 5.1 NAÏVE BAYES

Bayesian classifiers are statistical classifiers that predict the probability with which a given tuple belongs to a particular class[22]. Naïve Bayes classifier assumes that each attribute has its own individual effect on the class label, independent of the values of other attributes. This is called class-conditional independence. Bayesian classifiers are based on Bayes' theorem.

Bayes Theorem: Let X be a data tuple and C be a class label. Let X belongs to class C, then

$$P(C|X) = \frac{P(X|C)P(C)}{P(X)}$$

where;

- P(C|X) is the posterior probability of class C given predictor X.
- P(C) is the prior probability of class.
- P(X|C) is the posterior probability of X given the class C.
- P(X) is the prior probability of predictor.

The classifier calculates P(C|X) for every class Ci for a given tuple X. It will then predict that X belongs to the class having the highest posterior probability, conditioned on X. That is X belongs to class Ci if and only if

$$P(Ci|X) > P(Cj|X) \quad \text{for } 1 \leq j \leq m, j \neq i.$$

## 5.2 DECISION TREES

Decision tree induction is the process of creating decision trees for class-labeled training tuples[22]. A decision tree is basically a tree structure like a flowchart[22]. Each internal node of the tree represents a test on an attribute and each branch is the outcome of the test. Each leaf node is a class label. The first node at the top of the tree is the root node. To classify a given tuple X, the attributes of the tuple are tested against the decision tree starting from the root node to the leaf node which holds the class prediction of the tuple. Ross Quinlan introduced a decision tree algorithm called ID3 in his paper[23]. Later he introduced a successor of ID3 called C4.5 in[24] to overcome some shortcomings such as over-fitting. Unlike ID3, C4.5 can handle both continuous and discrete attributes, training data with missing values and attributes with differing costs. In a basic decision tree induction algorithm, all the training tuples are at the root node at start. The tuples are then partitioned recursively based on selected attributes. The attributes are selected based on an attribute selection method which specifies a heuristic procedure to determine the splitting criterion. The algorithm terminates if all the training tuples belong to the same class or there are no remaining attributes for further partitioning or all training tuples are used. ID3 uses the attribute selection measure called information gain, which is simply the difference of the information needed to classify a tuple and the information needed after the split. These two can be formularized as follows:

Expected information needed to classify a tuple in the training set D

$$Info(D) = -\sum_{i=1}^{m} p_i \, log_2(p_i)$$

where; pi is the nonzero probability that a tuple in D belongs to class Ci.
Information needed after the splitting (to arrive at the exact classification)

$$Info_A(D) = \sum_{j=1}^{v} \frac{|D_j|}{|D|} \times Info(D_j)$$

where A is the attribute on which the tuples are to be partitioned.
Then, information gain

$$Gain(A) = Info(D) - Info_A(D)$$

The attribute with highest information gain is selected as the splitting attribute.

C4.5 uses gain ratio as the attribute selection measure. Gain ratio is an extension to information gain in a sense because it normalizes information gain by using a split information value;

$$SplitInfo_A(D) = -\sum_{j=1}^{v} \frac{|D_j|}{|D|} \times log_2\left(\frac{|D_j|}{|D|}\right)$$

Then,

$$GainRatio(A) = \frac{Gain(A)}{SplitInfo_A(D)}$$

The attribute with the highest gain ratio is selected as the splitting attribute.

## 5.3 RANDOM FOREST

Random Forests is an ensemble method for classification and regression[22]. Random forests are a set of decision trees where each tree is dependent on a random vector sampled independently and with the same distribution of all the trees in the forest[25]. The algorithm generates a number of decision trees creating a forest. Each decision tree is generated by selecting random attributes at each node to determine the split[22]. Tim Kam Ho introduced the first method for random forests using random subspace method in his paper [26]. Later, Breiman Leo extended the algorithm in his paper [25] and this method was official known as Random Forests. The general procedure to generate decision trees for random forests starts with a dataset D of d tuples. To generate k decision trees from the dataset, for each iteration k, a training set Di of d tuples is sampled with replacement from the dataset D. To construct a decision tree classifier, at each node, a small number of attributes from the available attributes are selected randomly as candidates for the split at the node. Then Classification And Regression Trees (CART)[27] method is used to grow the trees. The trees are then grown to maximum size and are not pruned. CART is a non-parametric decision tree induction technique that can generate classification and regression trees. CART recursively selects rules based on variables' values to get the best split. It stops splitting when it detects that no further gain can be made or some pre-determined stopping conditions are met.

## 5.4 SUPPORT VECTOR MACHINE

Vladimir Vapnik, Bernhard Boser and Isabell Guyon introduced the concept of support vector machine in their paper [28]. SVMs are highly accurate and less prone to overfitting. SVMs can be used for both numeric prediction and classification. SVM transforms the original data into a higher dimension using a nonlinear mapping. It then searches for a linear optimal hyperplane in this new dimension separating the tuples of one class from another. With an appropriate mapping to a sufficiently high dimension, tuples from two classes can always be separated by a hyperplane. The algorithm finds this hyperplane using support vectors and margins defined by the support vectors. The support vectors found by the algorithm provide a compact description of the learned prediction model. A separating hyperplane can be written as:

$$W \cdot X + b = 0$$

where W is a weight vector, W = {w1, w2, w3,..., wn}, n is the number of attributes and b is a scalar often referred to as a bias. If we input two attributes A1 and A2, training tuples are 2-D, (e.g., X = (x1, x2)), where x1 and x2 are the values of attributes A1 and A2, respectively. Thus, any points above the separating hyperplane belong to Class A1:

$$W \cdot X + b > 0$$

and any points below the separating hyperplane belong to Class A2:

$$W \cdot X + b < 0$$

SVM was originally used for binary classification. However, several multiclass SVM algorithms have also been developed. In Weka, we used LIBSVM package developed by Chih-Chung Chang and Chih-Jen Lin[29]. The package can be downloaded from http://www.csie.ntu.edu.tw/ cjlin/libsvm. LIBSVM is an easy to use package to apply multiclass SVM and has gained a wide popularity in machine learning.

## 6. RESULTS AND DISCUSSION

We used different sizes of training and test sets to find the best combination that gives the most accuracy. We used four machine learning algorithms: Naïve Bayes, Decision Trees, Random Forest and Support Vector Machine in our experiments. The results are tabulated below. Table 1shows the accuracies of the algorithms for predicting runs and table 2 shows the accuracies of the algorithms for predicting wickets.

As we can see, Random Forest builds the most accurate prediction models for both batting and bowling in all the cases. Also, the accuracy of the models increases as we increase the size of the training dataset for all algorithms except in case of Naïve Bayes for batting where the accuracy decreases as we increase the size of the training set. Random Forest predicts runs with the highest accuracy of 90.74% when we use 90% of the dataset for training. Similarly, Random Forest predicts wickets with highest accuracy of 92.25% when we use 90% of the dataset for training. On the other hand, Naïve Bayes predicts runs with the least accuracy of 42.5% when we use 90% of the dataset for training. Naïve Bayes predicts wickets too with the least accuracy of 57.05% when we use 60% of the dataset for training. Decision Trees performs reasonably well with the maximum accuracy of 80.46% and the minimum accuracy of 77.93% for predicting runs. It predicts wickets with the maximum accuracy of 86.5% and the minimum accuracy of 84.4%, which is again reasonably well against the performance of Random Forest. The prediction models of SVM for predicting runs showed the maximum accuracy of 51.45% with 90% training data and the minimum accuracy of 50.54% with 60% training data. Also for wickets, SVM had the maximum accuracy of 68.78% with 90% training data and the minimum accuracy of 67.45% with 60% training data. Thus, surprisingly SVM was beaten by Random Forest and Decision Trees on both the datasets.

Table 1 Predicting runs

| Classifier | Accuracy (%) | | | |
| --- | --- | --- | --- | --- |
| | 60% train 40% test | 70% train 30% test | 80% train 20% test | 90% train 10% test |
| Naïve Bayes | 43.08 | 42.95 | 42.47 | 42.50 |
| Decision Trees | 77.93 | 79.02 | 79.38 | 80.46 |
| Random Forest | 89.92 | 90.27 | 90.67 | 90.74 |
| SVM | 50.54 | 50.85 | 50.88 | 51.45 |

Table 2 Predicting wickets

| Classifier | Accuracy (%) | | | |
|---|---|---|---|---|
| | 60% train 40% test | 70% train 30% test | 80% train 20% test | 90% train 10% test |
| Naïve Bayes | 57.05 | 57.18 | 57.48 | 58.12 |
| Decision Trees | 84.40 | 85.12 | 85.99 | 86.50 |
| Random Forest | 90.68 | 91.26 | 91.80 | 92.25 |
| SVM | 67.45 | 67.53 | 68.35 | 68.78 |

Muthuswamy and Lam[1] achieved an accuracy of 87.10% with backpropagation network (BPN) and 91.43% with radial basis function network (RBFN) table 3. They predicted wickets in two classes: 0 or 1 and 2 or more. Their study was limited only to eight Indian bowlers who played ODI matches for India since year 2000 against seven countries. Our approach works for any player in the world and is also applicable to new players who will be playing for their country in future. Moreover, they considered only three input parameters: An ID assigned manually to a bowler, an ID assigned manually to the opposition team and the number of overs bowled by the bowler. We carried out a detailed study of the stats and characteristics of the players by considering a number of attributes that can potentially impact the performance of a player in any match.

Table 3 Predicting wickets using BPN and RBFN[1]

| Model | Accuracy (%) |
|---|---|
| BPN | 87.10 |
| RBFN | 91.43 |

Table 4 summarizes the other performance measures of the algorithms with their best values for predicting runs and table 5 summarizes the other performance measures of the algorithms with their best values for predicting wickets.

Table 4 Performance measure of the algorithms for predicting runs

| Classifier | Precision | Recall | F1 Score | AUROC | RMSE |
|---|---|---|---|---|---|
| Naïve Bayes | 0.424 | 0.431 | 0.418 | 0.740 | 0.3808 |
| Decision Trees | 0.824 | 0.825 | 0.824 | 0.923 | 0.2409 |
| Random Forest | 0.908 | 0.908 | 0.908 | 0.987 | 0.1604 |
| SVM | 0.609 | 0.616 | 0.609 | 0.870 | 0.2908 |

As can be seen from the table, Random Forest performs the best in terms of all the measures woth precision, recall and F1 Score of 0.908, AUROC of 0.987 and root mean squared error of 0.1604 which are excellent values for a classifer. On the other hand, Naïve Bayes performs the worst with 0.424 precision, 0.431 recall, 0.418 F1 score, AUROC of 0.740 and root mean squared error of 0.3908. SVM also showed a poor performance with precsion of 0.609, recall of 0.616 and F1 score of 0.609 and root mean squared error of 0.2908. However, it AUROC value is good, which is 0.870. Decision Trees has performed well with precision and F1 score of 0.824, recall of 0.825, an excellent ROC value of 0.923 and RMSE value of 0.2409.

Table 5 Performance measure of the algorithms for predicting wickets

| Classifier | Precision | Recall | F1 Score | AUROC | RMSE |
|---|---|---|---|---|---|
| **Naïve Bayes** | 0.577 | 0.581 | 0.575 | 0.765 | 0.4216 |
| **Decision Trees** | 0.865 | 0.865 | 0.865 | 0.921 | 0.2812 |
| **Random Forest** | 0.923 | 0.923 | 0.923 | 0.975 | 0.2036 |
| **SVM** | 0.720 | 0.707 | 0.708 | 0.867 | 0.2905 |

Random Forest again performed the best for predicting wickets in terms of all the measures with precision, recall and F1 score of 0.923, AUROC value of 0.975 and root mean squared error of 0.2036. Again, Naïve Bayes shows the worst performance with 0.577 precision, 0.581 recall, 0.575 F1 score, 0.765 AUROC and root mean squared error of 0.4216. Decision Trees shows a good performance with precision, recall and F1 score of 0.865, AUROC of 0.921 and root mean squared error of 0.2812. SVM performed reasonably well with precision of 0.720, recall of 0.707, F1 score of 0.708, RMSE value of 0.2905but a good AUROC of 0.867.

# 7. CONCLUSION AND FUTURE WORK

Selection of the right players for each match plays a significant role in a team's victory. An accurate prediction of how many runs a batsman is likely to score and how many wickets a bowler is likely to take in a match will help the team management select best players for each match. In this paper, we modeled batting and bowling datasets based on players' stats and characteristics. Some other features that affect players' performance such as weather or the nature of the wicket could not be included in this study due to unavailability of data. Four multiclass classification algorithms were used and compared. Random Forest turned out to be the most accurate classifier for both the datasets with an accuracy of 90.74% for predicting runs scored by a batsman and 92.25% for predicting wickets taken by a bowler. Results of SVM were surprising as it achieved an accuracy of just 51.45% for predicting runs and 68.78% for predicting wickets.

Similar studies can be carried out for other formats of the game i.e. test cricket and T20 matches. The models for these formats can be shaped to reflect required characteristics of the players; e.g. batsmen need to have patience and ability to play longer innings in test matches whereas score more runs in less overs in T20 matches. Similarly, bowlers need to have stronger wicket taking abilities in test matches and better economy rate i.e. conceding less runs in T20 matches. Moreover, attempts can be made to improve accuracies of the classifiers for ODI matches.

## AUTHORS

Kalpdrum Passi received his Ph.D. in Parallel Numerical Algorithms from Indian Institute of Technology, Delhi, India in 1993. He is an Associate Professor, Department of Mathematics & Computer Science, at Laurentian University, Ontario, Canada. He has published many papers on Parallel Numerical Algorithms in international journals and conferences. He has collaborative work with faculty in Canada and US and the work was tested on the CRAY XMP's and CRAY YMP's. He transitioned his research to web technology, and more recently has been involved in machine learning and data mining applications in bioinformatics, social media and other data science areas. He obtained funding from NSERC and Laurentian University for his research. He is a member of the ACM and IEEE Computer Society.

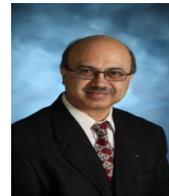

Niravkumar Pandey is pursuing M.Sc. in Computational Science at Laurentian University, Ontario, Canada. He received his Bachelor of Engineering degree from Gujarat Technological University, Gujarat, India. Data mining and machine learning are his primary areas of interest. He is also a cricket enthusiast and is studying applications of machine learning and data mining in cricket analytics for his M.Sc. thesis.

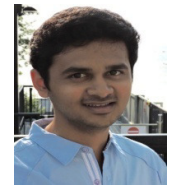